# Vaccination rates modulate the correlation between country latitude and covid-19 surge date arising this autumn 2021.

Stephan Walrand, Cliniques Universitaires Saint-Luc, 1200 Brussels, Belgium

stephan.walrand@uclouvain.be

**Abstract:**

A clear linear correlation of covid-19 surge date with country latitude was observed last autumn 2020, but not with temperature, nor with humidity, pointing the seasonal vitamin D status decline as a contributor factor. Well vaccinated European countries are this autumn 2021 repeating explosive surges. We aim to evaluate the potential link between surge date, country latitude and population vaccination rate.

Only a weak correlation was observed this autumn. However, after shifting the country latitude towards the south proportionally to its vaccination rate, a clear linear correlation was retrieved. This observation, joined to other ones, supports that vitamin D deficiency could reduce the vaccination efficiency. There is no objective benefit to hold a vitamin D deficiency. Thus, while European populations are undergoing the seasonal autumn-winter vitamin D status decline, it already makes sense to screen the elder people vitamin D status and fix individual deficiencies.

**Introduction:**

A recent randomized clinical trial reported a low vaccine efficacy around 75% in symptomatic COVID-19 middle east patients [1]. It was then advocated that this lower efficiency could be linked to vitamin D deficiency [2], the prevalence of which in this region being higher than in Europe [3]. This potential role of vitamin D is also supported by molecular mechanism analysis [4].

A clear linear correlation ($R^2$ = 0.77) of covid-19 surge date during autumn 2020 with population weighted latitude was observed, but not with temperature, nor with humidity, pointing the seasonal autumn-winter vitamin D status decline as a contributor factor [5]. Versus surge intensity analysis, surge date analysis has the benefit to be less sensitive to the differences in safety rules and in social uses existing between countries. After an increase in June, linked to the wide reduction of safety measures, European countries reached a more or less stable plateau.

European countries are this autumn 2021 repeating explosive covid-19 surges, even countries having a vaccination rate above 65%. The aim of this report is to evaluate the potential link between surge date, country latitude and population vaccination rate.

**Methods:**

All data used are public and were extracted from:

- the John Hopkins University covid-19 data (https://github.com/CSSEGISandData/COVID-19) via the Google interactive page for the daily positive cases allowing the autumn surge date determination (see .pptx in the supplementary files)

- from the Baylor University population resource (http://cs.ecs.baylor.edu/~hamerly/software/europe_population_weighted_centers.txt) for the population weighted latitude

- from Our World in Data (https://ourworldindata.org/covid-vaccinations?country=OWID_WRL) for the country vaccination rates (see .pdf for the vaccination rates assessed at October 2021)

The latitude shift providing the optimal surge date correlation was obtained using the Excel solver (see .xlsx in the supplementary files)

**Results:**

Table 1 shows the country data:

| country | surge date | vac. % | latitude [deg] | shifted lat. [deg] |
|---|---|---|---|---|
| Portugal | oct 25 | 85 | 39.7 | 14.4 |
| Spain | oct 29 | 78 | 39.8 | 16.6 |
| Greece | oct 06 | 59 | 38.7 | 21.1 |
| Bulgaria | sept 25 | 20 | 42.8 | 36.8 |
| Italy | oct 21 | 71 | 42.9 | 21.7 |
| Serbia | oct 14 | 43 | 43.8 | 31.0 |
| Croatia | oct 12 | 43 | 45.3 | 32.5 |
| Slovenia | oct 11 | 51 | 46.2 | 31.0 |
| France | oct 09 | 68 | 47.2 | 26.9 |
| Austria | oct 11 | 62 | 47.8 | 29.3 |
| Belgium | oct 09 | 74 | 50.8 | 28.7 |
| Germany | oct 12 | 66 | 50.9 | 31.2 |
| Netherland | oct 03 | 67 | 52.1 | 32.1 |
| UK | sept 18 | 68 | 52.8 | 32.5 |
| Russia | sept 10 | 32 | 54.3 | 44.8 |
| Finland | sept 22 | 67 | 61.8 | 41.8 |
| Denmark | sept 20 | 75 | 55.9 | 33.5 |
| Ireland | oct 07 | 75 | 53.1 | 30.7 |
| Ukraine | sept 02 | 15 | 48.8 | 44.3 |

Figure 1 shows that a weaker linear correlation ($R^2 = 0.22$) with the country latitude is now observed (orange bullets). However, when the country latitude is shifted proportionally to its population vaccination percentage (black arrows), a clear linear correlation is retrieved ($R^2 = 0.66$, blue bullets). In order words, vaccination rates virtually shift the country towards the equator.

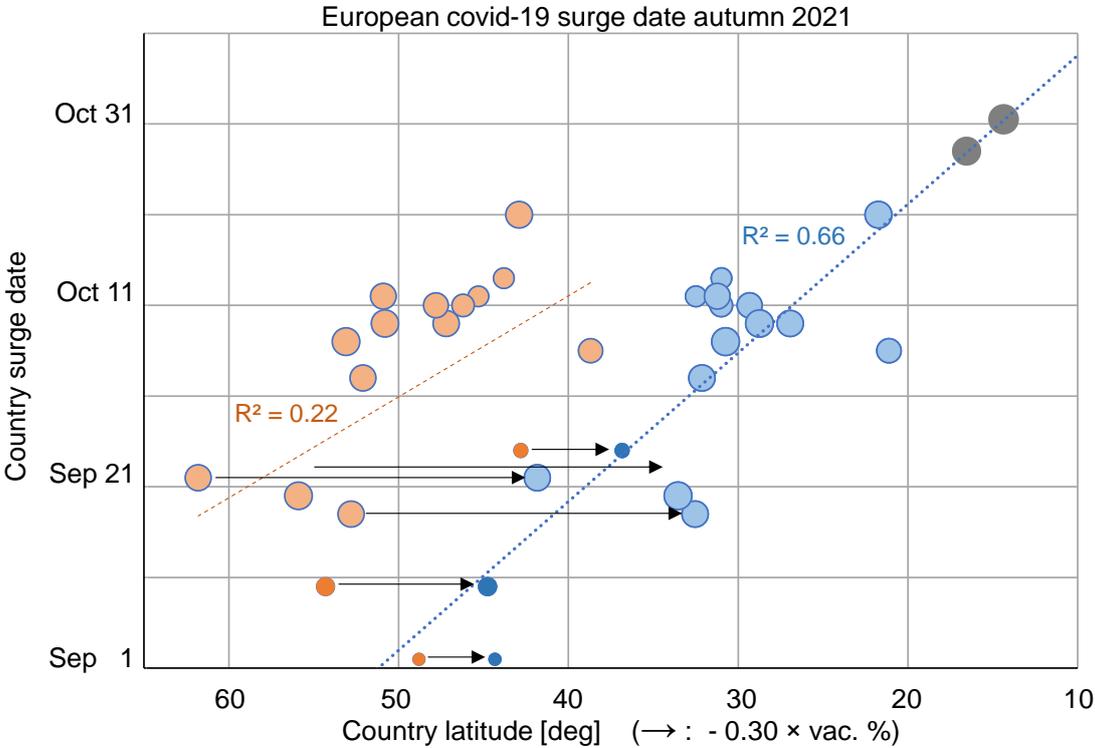

Figure 1: country surge date as a function of the population weighted latitude (orange bullets) and shifted (black arrows) proportionally to the country vaccination rate (blue bullets). The bullet diameter is proportional to the country vaccination rate. Note the low shift for the three countries having the lowest vaccination rates (dark orange: Ukraine 15%, Bulgaria 20%, Russia 32%) versus the other ones. Two southern countries having the highest vaccination rates (grey bullets: Portugal 85%, Spain 78%) have not yet undergone a surge and have been represented at their expected surge dates (obviously, they have not been included in the computation of $R^2$). It is possible that these 2 southern and highly vaccinated countries will never face an autumnal surge.

**Discussion:**

This analysis confirms the correlation of the surge date with the latitude observed in autumn 2020 [4] and clearly supports that a high vaccination rate has a significant dumping effect on the virus transmission, despite that the delta variant is now common in all European countries. A contrario, it also supports that vitamin D deficiency could reduce the vaccination efficiency.

**Conclusion:**

Furthers investigation about the link between lower vaccination efficiency and vitamin D deficiency are suitable. However, there is no benefit to hold a vitamin D deficiency. Thus, while European populations are undergoing the seasonal autumn-winter vitamin D status decline, it already makes sense to screen the vitamin D status of elder populations in whom deficiencies are more frequent and fix individual deficiencies.

**Author ontributions:**

SW is the sole contributor of this article.

**Competing interests:**

SW declares to have non conflict of interest.

**Supplementary information** is available for this paper

Correspondence and requests for materials should be addressed to stephan.walrand@uclouvain.be

**References:**

(1) Al Kaabi N, Zhang Y, Xia S, Yang Y, Al Qahtani MM, Abdulrazzaq N, et al. Effect of 2 inactivated SARS-CoV-2 vaccines on symptomatic COVID-19 infection in adults: a randomized clinical trial. JAMA 2021;326:35–45.

(2) Wu CC, Lu KC. Vitamin D deficiency and inactivated SARS-CoV-2 vaccines. Eur J Int Med. 2021;93;114-114.

(3) van Schoor N, Lips P. Global overview of vitamin D status. Endocrinol Metab Clin North Am 2017;46:845–70.

(4) Chiu SK, Tsai KW, Wu CC, Zheng CM, Yang CH, Hu WC, Hou YC, Lu KC, Chao YC. Putative Role of Vitamin D for COVID-19 Vaccination. Int J Mol Sci. 2021;22:8988.

(5) Walrand S. Autumn COVID-19 surge dates in Europe correlated to latitudes, not to temperature-humidity, pointing to vitamin D as contributing factor. Sci Rep. 2021;11:1-9.